\documentclass{JHEP}
\usepackage{amssymb} 
\usepackage{amsfonts}
\newcommand{\beq}{\begin{equation}}
\newcommand{\eeq}{\end{equation}}
\newcommand{\beqs}{\begin{eqnarray}}
\newcommand{\eeqs}{\end{eqnarray}}

\newcommand{\Tr}{{\mathrm{Tr}}}
\newcommand{\dd}{{\mathrm d}}
\newcommand{\ee}{{\mathrm e}}
\newcommand{\rr}{\mathrm{RR}}
\newcommand{\ns}{\mathrm{NS}}
\newcommand{\cp}{\mathrm{CP}}
\newcommand{\ZZ}{{\mathbb Z}}

\newcommand{\RR}{{\mathbb R}}
\newcommand{\CC}{{\mathbb C}}
\newcommand{\FF}{{\mathbb F}}
\newcommand{\Cech}{{\v{C}ech~}}
\newcommand{\half}{\frac{1}{2}}

\newcommand{\vol}{{\mathrm{vol}}}
\newcommand{\Vol}{{\mathrm {Vol}}}

\title{Gerbes and Massive Type II Configurations}

\author{
Jussi Kalkkinen\thanks{This work was supported in part 
by the European Union TMR program CT960045.} \\ 
SISSA, via Beirut 4, Trieste 34014, Italy \\
INFN, Sezione di Trieste \\ 
E-mail: {\tt kalkkine@sissa.it}
}

\abstract{We find novel bound states of NS5, D6 and D8-branes in massive 
type IIA string theory. 
As the NS gauge transformations can change the Chern 
class of the RR field these configurations should be thought of as 
nonlocal objects called gerbes. 
We develop a global formalism for theories that involve 
massive tensor fields
in general, and apply it in massive type IIA supergravity. 
We investigate the T-dual 
NS5/D5-brane   configurations in type IIB, and relate
them to an F-theory compactification on Calabi--Yau three-fold. 
We comment on the implications to consistency conditions 
for brane-wrapping and the classification of 
D-brane charges in terms of K-theory classes.}


\preprint{SISSA 46/99/EP \\ 
\hepth{9905018}}

\begin{document}

\email{kalkkine@sissa.it}

\section{Introduction}

Gauge symmetry has certainly proved to be one of
the most efficient tools both in quantum field theory and  
in string theory. Physically it  
facilitates the study of complicated systems by 
enlarging the space where these systems are embedded,
and ensures the stability of, for instance, 
such solitonic objects as instantons.
These physical ideas are usually formulated mathematically
in the language of fiber bundles, which 
are global constructions that encode much 
relatively easily accessible information of the 
topology of the underlying manifold.
In this article we give examples of string theory 
configurations that go beyond the
realm of globally defined bundles and belong 
in a sense to a stringy extension of gauge theories. 

In a well defined theory two symmetries
might seem to conflict in that
one of them could shift the conserved charge of the other.
This situation arises for instance 
in any theory that involves massive tensor fields, such as many 
world-volume effective theories in string theory, or the massive 
type IIA supergravity. There a gauge symmetry of one field seems 
to shift the characteristic class of another field. This situation 
seems to fall outside the usual mathematical discussion
that makes use of everywhere defined bundles.
Recently, there have appeared 
constructions that seem to be able to accommodate this kind of 
situations \cite{Brylinski, Hitchin}. 
These constructions, {\em gerbes}, are most conveniently
described in terms of {\em locally} defined line bundles, as 
opposed to the everywhere defined fiber bundles we referred 
to above. They are the next logical step in a sequence 
that starts with functions on a manifold, and is 
followed by line bundles. These constructions 
have already been applied in quantum field theory in 
\cite{Mickelsson:1995cr, Carey:1995wu, Carey:1997xm}.

In this paper we present some inherently nonlocal objects
in string theory, and show how to analyze them as gerbes.
These objects arise naturally when one
employs the symmetries of the effective type IIA supergravity 
theory to the full. The example at hand is 
a bound state of NS5-branes and 
D6-branes in the presence of a cosmological constant. 
Gerbes cannot be seen as local objects in the same sense as 
normal bundles or Chan--Paton bundles can. This fact 
also sheds light on 
some of the difficulties encountered in classifying
type IIA D-branes in terms of fiber bundles.

In \cite{Minasian:1997mm} it was suggested that D-brane charges 
could be described by K-theory classes. The classification of 
a large class of the cases encountered 
in string theory was performed (for trivial NS two-form backgrounds) 
in \cite{Witten:1998cd}, and continued in 
\cite{Horava:1998jy, Garcia-Compean:1998rg, Gukov:1999yn}. 
The classification was based on the idea that supersymmetric 
D-branes  could be built as bound states of non-supersymmetric 
higher-dimensional D-branes \cite{Sen:1998sm}. 
The  stability of a these
brane--anti-brane 
constructions depends on the properties of the branes' normal and
Chan--Paton bundles. 
On the other hand, vector bundles 
on a manifold $X$ are in general classified in K-theory. 
This leads to a partial 
classification of D-brane charges in terms of the groups $K(X)$ in type IIB 
and $K^1(X)$ in type IIA.
The classification was also extended to comprise the so-called 
twisted fiber bundles, which are essentially trivial gerbes. 
Having introduced gerbes as generalizations of fiber bundles the 
next logical step would be to extend this classification 
to something that should then be the ``K-theory of gerbes'', 
in order to cope with general NS two-form backgrounds and the 
cosmological constant in the case of type IIA.
This still remains to be done.

In all of these configurations the integral cohomology 
class of the NS field strength plays a central role. It 
appears as an obstruction to wrapping a brane on a 
submanifold or, in certain cases, to classifying D-brane 
charges in K-theory, {\em and} it is the characteristic 
class of a gerbe. These appearances are intimately connected.
In this paper we attempt among other things to clarify the 
role played by the two-form NS field  in string theory.
In the same way that a line bundle has a natural 
characteristic class related to, say, in the case 
of the Dirac monopole bundle to the number of 
monopoles, the characteristic class of the gerbe 
will in our constructions describe the number of NS5-branes. 
The local line bundles needed for defining a gerbe 
imply that there are, in a certain sense, nonlocal 
D6-branes. Nevertheless, we can find a globally defined 
modified eight-form field strength. The topological 
stability of these constructions follows from 
the gerbe structure.

Another  central ingredient will be the zero-form 
field strength in type IIA string theory, the cosmological 
constant. We shall consider it a section of a constant sheaf; 
Allowing it to jump would indicate the presence of a D8-brane. For 
the most part of the paper the cosmological constant will be 
a true, though non-zero, constant as turns out to be necessary 
for the consistency of a gerbe.
String theory would seem to resolve, as we
shall argue, at least those problems
that arise from the presence of the cosmological constant 
by simply relating the problematic type IIA solutions 
to more classifiable ones in type IIB. 

In order to gain further understanding of these systems 
we find
the T-dual solutions on the type IIB side. These turn out to 
include 
a non-trivial axion, NS5/D5-brane dyons, D7-branes and orbifold 
singularities. 
Stability considerations in IIA lead one 
to a consistent treatment of the deficit angle in type IIB. The 
form-field equations of motion can be solved explicitly, but the 
metric 
and the dilaton must be extracted from an F-theory construction 
on an 
elliptically fibered Calabi--Yau three-fold.

The article is organized as follows: We begin with a toy 
model that involves massive tensor fields in general.
We then give a brief account of what a gerbe is. We 
present our ansatz in massive type IIA and 
solve the form-field equations of motion, 
dualize the solutions, and relate the construction 
to an F-theory solution. We then discuss the classification 
of in particular type IIA branes in K-theory and the consistency 
conditions for brane wrapping in the last section.

\section{Global structures in field theory}

\subsection{Massive form-fields}
\label{tensorfields}
The constructions that we shall make use of below are characteristic 
to any theory, where tensor fields become massive. These theories 
include the Maxwell--Proca theory, Romans' massive IIA supergravity 
\cite{romans}, 
and effective world-volume theories on branes  
\cite{Strominger:1996ac, Townsend:1996em}.  
In order to extract the essential features of these systems, it may 
be helpful to start by considering the toy model
\beqs
S = \int_X \dd C \wedge *  \dd C  + ( m C + \dd L) \wedge * ( 
m C + \dd L) 
+ \ldots~. 
\eeqs 
Here $C$ is a $(p+1)$-form field over a space-time $X$, 
$L$ is a $p$-form sometimes 
called the St\"uckelberg field, and $m$ is a constant. 
The action is invariant under the transformations
\beqs
C &\longrightarrow& C + \dd A \nonumber \\
L &\longrightarrow& L -  m A \label{g1}
\eeqs
and
\beqs
C &\longrightarrow& C \nonumber   \\
L &\longrightarrow& L + \dd a~. \label{g2}
\eeqs
In general these transformations remain symmetries of the theory, 
when e.g.~$\dd A$ is replaced by a closed form, say $F$. 
In this case, however, $A$  is defined only locally. 

In order to see what physical states there are one has  to fix 
the gauge. Any 
configuration $(C,L)$ lies locally on the gauge orbit of 
$(C+\dd L, 0)$. The $L$-field is thus not physical, 
and can be gauged away. In this gauge $C$ is obviously a massive 
tensor field, the mass being given by the parameter $m$.
One may ask what happens in the theory when one takes the  
limit $m \rightarrow 0$. 
The two fields decouple, and the physics is described by a 
massless tensor field
and a $U(1)$ gauge theory.
The physical Hilbert spaces of the theory are rather
different depending on the value of $m$.

\subsection{Transition functions}

Physical states can generally be described as 
equivalence classes of 
fields modulo symmetries. When we are building a theory on a 
topologically non-trivial space we must first cover the space 
(say, $X$) by a suitable collection of local charts 
$\{ {\cal U}_{\alpha} \}$. 
Fields defined on different charts are related by symmetries of 
the theory -- this means simply that the actions of transition 
functions 
on different fields are given essentially by gauge transformations. 
In this well known way e.g.~tensor 
fields become sections of $\Omega^*(X,\RR)$.

A particular configuration is described so
that one gives the (not yet gauge fixed) fields on each 
coordinate patch, and tells what gauge transformation 
$g_{\alpha\beta}$ 
has to be done on the intersection  ${\cal U}_{\alpha} 
\cap {\cal U}_{\beta}$ to interpolate between different 
local descriptions. For instance, in the example 
above we can take $C \in \Omega^{p+1}(X,\RR)$ and set
\beqs
C_{\beta} = g_{\beta\alpha} \cdot C_{\alpha}  
= C_{\alpha} + \dd A_{\alpha\beta}~.
\eeqs

From this construction certain consistency conditions ensue. 
For example the transition functions of a vector bundle 
must obey the cocycle condition 
\beqs
(\delta g) = g_{\alpha\beta}g_{\beta\gamma}g_{\gamma\alpha} = 1 
\label{2-cocycle}
\eeqs
on triple intersections of local charts.
Gluing the local theories together on a given space is not 
always possible; obstructions to this are most conveniently 
discussed 
in \Cech cohomology, cf.~Section \ref{cechsec}.

Let us apply this to the toy model of Section \ref{tensorfields}:
Given an atlas ${\cal U}_\alpha$ on the space $X$  
we can define a field theory where fields 
on different patches are related by the gauge transformations 
(\ref{g1}) and (\ref{g2}).
The latter transformations 
suggest for $p=1$ an underlying 
line bundle  ${\cal L} \rightarrow X$,  
provided the global extension of the local form  $\dd L$ 
defines an integral cohomology class. 
This class characterizes the topology of the bundle.
The former gauge transformations have a similar action on $C$. 
These transformations, however, have a  
peculiar action on the line bundle ${\cal L}$; it seems 
as if we could shift the characteristic class of 
the bundle ${\cal L}$ {\em by a symmetry of the theory}. 
This suggests to consider structures where there are different 
line bundles on different coordinate patches. These 
structures are most conveniently described in terms of gerbes.

\section{Gerbes}
\label{gerbe}

As was in the case of the ``ambiguities'' introduced by gauge 
symmetries, even the ambiguity in defining the right line bundles 
for 
St\"uckelberg fields can be turned into an asset. 
In the same way that the former lead us to work with 
fiber bundles, the latter should lead us to consider gerbes. 
Here we shall follow the lecture notes of Hitchin \cite{Hitchin}.

\subsection{\Cech cohomology}
\label{cechsec}

Let $\FF=\ZZ_2$ or $U(1)$ and assign to each $n$-fold intersection 
of local charts a map to $\FF$ 
\beqs
f_{i_1,\cdots,i_n} : 
U_{i_1} \cap \cdots U_{i_n} \longrightarrow \FF~. \label{chain}
\eeqs
that is antisymmetric in its indices.
Then we can define a coboundary operator by setting
\beqs
(\delta f)_{i_1,\cdots,i_{n+1}} = \exp \sum_{k=1}^{n+1} (-1)^{k+1} \log 
f_{i_1,\cdots,\hat{k},\cdots,i_{n+1}}~.
\eeqs
For example, the transition functions of an orthogonal or a 
unitary vector bundle define such functions for $n=2$, namely 
$f_{\alpha\beta} = \det(g_{\alpha\beta})$. 

The cohomology $\check H^*(X,\FF)$  of this complex, namely the 
cocycles
(i.e.~functions that satisfy $\delta f =0$) modulo coboundaries 
(i.e.~exact functions $f =\delta h$),  
is called \Cech cohomology. 
It characterizes the topology of the manifold and not only the 
choice of cover, if the cover 
is fine enough \cite{Spanier}. 
For instance, the first Stiefel--Whitney class  
$w_1 \in \check H^1(X,\ZZ_2)$ which is constructed using transition 
functions of a tangent bundle 
is an obstruction to orientability. The second  
Stiefel--Whitney class $w_2$ similarly over $\ZZ_2$ is 
constructed using lifts of these transition functions 
into the spin-cover. A non-trivial $w_2$ obstructs spin-structure.

\subsection{Definitions}

A space $X$ together with a collection of maps 
$g_{\alpha\beta\gamma}$ as in (\ref{chain}) for $n=3$
is called a gerbe \cite{Hitchin} if
\beqs
(\delta g)_{\alpha\beta\gamma\delta} = 1~.
\eeqs
This is an obvious analogue of the condition (\ref{2-cocycle}). 
In the same way that a Chern class characterizes a line bundle, 
a class in 
$\check H^3(X,\ZZ)$ characterizes a gerbe. 

There is an equivalent but a somewhat more concrete 
definition of a gerbe that makes use of local line bundles 
\cite{Hitchin}. It 
postulates that there exist
\begin{itemize}
\item[1)] A line bundle ${\cal L}_{\alpha\beta}$ on each 
intersection 
${\cal U}_{\alpha}  \cap {\cal U}_{\beta} $.
\item[2)] An isomorphism ${\cal L}_{\alpha\beta} \simeq 
{\cal L}_{\beta\alpha}^{-1}$. 
\item[3)] A trivialization $\theta_{\alpha\beta\gamma}$ of
\beqs
{\cal L}_{\alpha\beta} \otimes {\cal L}_{\beta\gamma}  
\otimes {\cal L}_{\gamma\alpha} \label{trivbundle}
\eeqs
that satisfies 
$(\delta \theta)_{\alpha\beta\gamma\delta} = 1$.
\end{itemize}
The local trivialization of a gerbe on a
two-fold intersection of local charts ${\cal L}_{\alpha\beta}$ 
is a 
{\em difference of line bundles},
\beqs
 {\cal L}_{\alpha\beta} = {\cal L}_{\alpha} \otimes 
{\cal L}_{\beta}^{-1}~.
\eeqs
This should be seen as an analogue of the observation that 
a difference (quotient)
of two trivializations of a line bundle (i.e.~non-zero sections) 
is an ordinary function.
In general it is possible to define higher gerbes as differences 
of lower gerbes; here
we shall only make use of the one-gerbes defined above.

\subsection{Connection}

Similarly to the curvature two-form of a line 
bundle that can be used to represent the first Chern 
class, there is a three-form that can represent the 
characteristic class of a gerbe. 
The three-form is constructed using
a locally defined two-form gauge potential 
$B_{\alpha} \in \Omega^2({\cal U}_{\alpha})$. 
The forms must satisfy the compatibility condition that 
the difference on a two-fold intersection 
${\cal U}_{\alpha} \cap  {\cal U}_{\beta}$ 
\beqs
B_{\alpha}  -  B_{\beta} = F_{\alpha\beta}
\eeqs
be a curvature of a {\em local line bundle}. 
The curvature of the gerbe is then simply
a global three-form  $[G] = \dd B_{\alpha} \in H^3(X,\RR)$. 
Similarly to the Chern class of a line bundle, 
also the curvature of a gerbe 
must be in integral cohomology.

This is exactly the situation we encountered 
in the massive tensor theories: a local two-form 
that can shift the characteristic class 
of a local line bundle 
by a symmetry of the theory.

\section{An ansatz in massive type IIA}

Type IIA string theory provides a physical 
example of a system, where the considerations of the previous 
section 
are realized.
There the mass parameter on the level of supergravity 
\cite{romans} 
arises as  the vacuum expectation value of the dual of 
a top-form field 
strength, cf.~\cite{polchinski, Bergshoeff:1996ui}. Furthermore, 
{\em both} of the 
equivalents of the fields $C$ and $L$ turn out to carry
charges associated to nonperturbative excitations in the full 
string 
theory, namely NS5-branes and D6-branes. 
It is therefore necessary to understand the spectrum in the 
$m\neq0$ phase 
without going directly to the naive unitary gauge $L=0$.

Below we shall present an ansatz, and show that it is 
consistent with all Bianchi identities and   
form-field equations of motion. 
In Section  \ref{metric} we relate the T-dual of the 
ansatz to a system of fivebrane dyons in F-theory 
compactified on a Calabi--Yau three-fold.

\subsection{The modified field strength}

Let $C$ be the free sum of RR fields. 
The modified field strengths of type II field theories are 
\cite{Green:1996bh}
\beqs
R(C) = \dd C - G \wedge C +m \ee^B~,
\eeqs
where $B$ is the NS two-form, $G = \dd B$, and $m$ the mass 
parameter of the type IIA theory. 
The Bianchi identities are
\beqs
\dd R(C) - G \wedge R(C) = 0 \label{rc}~.
\eeqs
The equations of motion follow 
from Romans' action \cite{romans}, which    
in our notation is
\beqs
S &=& \int  \ee^{-2 \phi} \big( *R 
 + 4 \dd \phi \wedge *\dd \phi  - \half  G \wedge *G \big) 
\nonumber  \\
& & - \half R(C) \wedge *R(C) + \half \dd^{-1} \big(  R^{[4]} 
\wedge  
R^{[4]} \wedge G \big) \label{full}
\eeqs
the only non-trivial components of $R^{[2k]}$ being those for $2k=0,2,4$.
Dualizing in the path integral leads to the dual field strengths 
$R^{[10-2k]} = (-)^{k} *R^{[2k]}$ as in \cite{Janssen:1999sa}.

Let $A_{\cp}$ be the Chan--Paton gauge potential supported on the 
brane world-volumes. Then 
the RR form symmetry acts on the various fields 
according to \cite{Green:1996bh}
\beqs
\delta_{\rr} C &=& \dd \Lambda - G \wedge \Lambda + \lambda 
\ee^B \nonumber\\
\delta_{\rr} B &=& 0 \label{rr} \\
\delta_{\rr} A_{\cp} &=&0~.   \nonumber 
\eeqs
The system possesses also a NS symmetry, namely
\beqs
\delta_{\ns} C &=& - m  \eta~ \ee^B \nonumber \\
\delta_{\ns} B &=& \dd \eta \label{ns} \\ 
\delta_{\ns} A_{\cp} &=& 2\pi \eta~. \nonumber
\eeqs
As explained above, when we build a twisted theory 
on a manifold it does not matter what the exact values 
of $C$ and $B$ are. Instead, the physical condition is 
that the modified field strengths $R(C)$ and $G$
be globally defined sections of $\Omega^*(X)$.
The RR symmetry is 
the ordinary form symmetry in disguise. 
We can, again, allow 
$\dd \eta \longrightarrow f$ to 
be cohomologously non-trivial. Then the NS symmetry, for 
$m\neq0$, 
changes the Chern classes or intersection numbers of 
the RR fields.

\subsection{Topology and RR fields}
\label{ansatz}

We consider an ansatz that supports NS5-brane, D6-brane 
and possibly D8-brane charges in massive 
type IIA theory. These
configurations including their supersymmetries 
have been discussed in a
different context for example in \cite{Hanany:1997ie, 
Hanany:1997sa}. 
Many non-supersymmetric solutions 
of the massive type IIA supergravity were 
constructed already in Ref.~\cite{romans}.
Supersymmetric solutions 
include the D8-brane of \cite{Polchinski:1996df} and 
\cite{Bergshoeff:1996ui}. Other solutions were found in 
\cite{Janssen:1999sa, Massar:1999sb}.

In order to find the relevant perturbative Lagrangian we first 
consider the Bianchi identities (\ref{rc}). 
It will turn out to be consistent to set 
$C^{[3]} = C^{[5]} = 0$ and to assume $B \wedge B =0$. 
From the duality equations it then 
follows that
\beqs
\dd C^{[7]} &=& -* (\dd C^{[1]} +m B)  \label{a1} \\
\dd C^{[9]} - G \wedge C^{[7]}  &=& * m~. \label{a2}
\eeqs
From (\ref{a1}), (\ref{a2}) and the fact 
that we shall be able to choose a $B$-field that satisfies
$B \wedge B =0$ it follows that it is sufficient to consider 
the action
\beqs
S_0 &=& \int  \ee^{-2 \phi} \big( *R 
 + 4 \dd \phi \wedge *\dd \phi  - \half  G \wedge *G \big) 
\nonumber  \\
& & - \half R^{[8]} \wedge *R^{[8]} + \int C^{[7]} \wedge 
\tau  \label{action}~,
\eeqs
where $\tau$ is the Poincar\'e dual of the D6-brane. 
Also the anomalous couplings
should be included, but it turns out that the equations of motion 
remain the same provided
$ \Tr( F \wedge F \wedge F)   = 0$, where $F =\dd A_{\cp}$.

Consider\footnote{The present ansatz makes use of a construction 
in \cite{Hitchin}.}
a space-time locally of the form 
$M^{1,5} \times \RR \times S^3$, and a fixed point $p \in S^3$.  
We cover the  sphere with two hemispheres ${\cal U}_0$ and 
${\cal U}_1$, so that the 
point $p$ is outside ${\cal U}_0$.
Let $H$  satisfy
\beqs
\Delta H = [V] - 2\pi  \delta_p \in H^3(S^3, \ZZ)~,
\eeqs
where $[V]$ is the de Rham class of the volume form 
i.e.~generator of $ H^3(S^3, \ZZ)$, 
$\delta_p$ is the Poincar\'e dual of the point $p$, $Q$  
is a parameter, 
and 
$\Delta = \{ \dd, \dd^\dagger \}$ is the Hodge--de Rham 
Laplacian\footnote{The Hodge star $*$ and the operator 
$\dd^\dagger$ refer to operations on $S^3$ and to the 
 metric normalized to $\int *1 =2\pi$. \label{hosta}}. 
On the coordinate patch ${\cal U}_1$ we can similarly define 
\beqs
\Delta H_1 = [V]~.
\eeqs 
For finding the three-forms $H,H_1$ it was necessary that the 
relevant cohomology classes on the right-hand side vanish. 

We can now give our ansatz for the various form-fields.
The NS two-form is locally defined 
\beqs
B = \left\{  
\begin{tabular}{ll}
$Q~ \dd^\dagger H$ &\, on\, ${\cal U}_0$ \\ 
$Q~ \dd^\dagger H_1$ &\, on \, ${\cal U}_1$
\end{tabular}
 \right. ~.
\eeqs
The field strength 
\beqs
G = \dd B = Q~[V]
\eeqs
is globally defined.
This ansatz supports $Q$ NS5-branes that are cut out of 
the  origin in  
$\RR^4 \backslash \{0\} \simeq \RR \times S^3$.

For generality we consider the mass parameter in Romans' 
supergravity as a locally constant $\RR$-sheaf
\beqs
m = \left\{  
\begin{tabular}{ll}
$m_0$ & \, on \, ${\cal U}_0$ \\ $m_1$ &\, on \, ${\cal U}_1$
\end{tabular}
 \right. ~.
\eeqs
This may lead to D8-brane charge. Before introducing the
metric the D8-brane is 
not exactly localized. 
On the intersection of charts  ${\cal U}_0 \cap {\cal U}_1$ 
the charge 
is the difference in cosmological constants 
$q(D8) = m_1 - m_0$.

As we are interested in bound states that involve D6-branes 
the correct electric RR-form is  $C^{[7]}$, whereas $C^{[1]}$ 
should be regarded as its magnetic dual. These are naturally 
both gauge dependent objects, and the field that we should 
match over the intersections is $R^{[8]}$.
It is convenient to consider the difference
\beqs
f = \Big(\dd C^{[1]}\Big)_1 - \Big(\dd C^{[1]}\Big)_0 = 
 m_0 B_0 - m_1 B_1
\eeqs 
and to assume
e.g.~$(\dd C^{[1]})_0=0$. 
The D6-brane charge can be calculated integrating over 
the charge density on a disc $D \subset {\cal U}_1$
\beqs
\int_{D} \dd f &=& \int_{D} (m_0-m_1)~Q~ V - 
2\pi m_0 Q~\delta_p  \\
&=& (m_0-m_1)~Q~ \vol(S^2)~  -~ 2\pi m_0 Q~. \label{fl}
\eeqs
Let us now investigate the various field strengths in detail:
Considering $m_0=m_1=m$ 
we only see the contribution coming from the extracted point.
The physical interpretation is that in that point a Dirac 
string emerges through the sphere 
$S^3 \subset \RR^4$ -- this is just the D6-brane.

In order to see what happens if $m_0 \neq m_1$ we 
should modify the above calculation 
by setting $\Delta H = [V]$, which is possible locally on 
${\cal U}_0$. The total NS field 
strength is  globally defined; physically this means that 
we can with impunity put NS5-branes inside a D8-brane. 
We can stretch the disc $D$ over which integration is performed
as to cover all of the sphere but a point. 
In this limit the integral yields
\beqs
\int_{\RR^3} \dd f 
&=& 2\pi~ (m_0-m_1)~Q~.  \label{flcn}
\eeqs
This means that 
there is a flux  corresponding to $(m_0-m_1)Q$ D6-branes  
at the excluded point. 
A similar phenomenon was observed in \cite{Hanany:1997sa}, where 
passing a NS5-brane through a D8-brane produced a 
D6-brane tube between the  NS5 and D8-branes.

The two-fold intersection of charts
${\cal U}_0 \cap {\cal U}_1$ is topologically a cylinder. We shall 
find it convenient to describe it as
a line bundle $\pi: C \longrightarrow S^2$ over a two-sphere. 
As $f$ is closed outside $p$ we can write for concreteness 
\beqs
f &=& \frac{m Q}{2}~ \dd \cos \theta \wedge \dd \varphi~, 
\label{wrap}
\eeqs
where $(r, \theta, \varphi)$ are the local coordinates and 
we are sufficiently near $p$. 
As the formula does not depend on $r$ it 
descends naturally to a two-form on the sphere $S^2$. 
Further away from $p$ the formula depends on the details of
the fibration of $C$.
If we can show that $f$ is in integral cohomology we have 
constructed a gerbe. We defer this discussion to Section 
\ref{quant}.
The flux $f$ is not defined far a way from $p$, but it is 
purely a property of the cylinder intersection of the 
two charts. 
The gauge invariant physical quantity is
$R^{[8]} = -* R^{[2]}$. On ${\cal U}_0$ we have 
$R^{[2]} = m B_0$. This actually extends to the whole 
sphere as $R^{[2]} = m Q \dd^\dagger H$. 
The electric field strength is 
\beqs
R^{[8]} = \frac{mQ}{b}~ \dd h \label{r8} \wedge {\Vol}(M^{1,5} 
\times \RR),
\eeqs
where $b$ is the conformal factor in front of the sphere metric 
in ten dimensions.
The function $h$ satisfies
\beqs
\nabla^2 h &=& 1 - 2\pi \delta^{(3)}(p)~.
\eeqs
The equation of motion is satisfied provided we insert 
the following six-brane source
\beqs
\tau= mQ~(V - 2\pi~ \delta_p)~.
\eeqs
The D6-branes are supported in $p$ {\em and} spread evenly 
over the sphere. 
The NS two-form equation of motion reduces to ensuring that  
\beqs
\ee^{2\phi}~ \frac{\vol(M^{1,5} \times \RR)}{\vol(S^3)}
\eeqs
does not depend on the coordinates of the sphere.

\subsection{Charge quantization}
\label{quant}

We want to interpret the non-gauge invariant two-form $f$ 
as the curvature of a suitable line bundle on the cylinder 
$C: {\cal U}_0 \cap {\cal U}_1 \simeq S^2 \times \RR$. If 
this can be done
the whole of the structure extends to a gerbe, as 
was explained  in Section \ref{gerbe}.
The bundle we are looking for ${\cal L}_{01} \longrightarrow 
S^2$ 
can be taken to have the same base space $S^2$ as the cylinder 
itself. 
The condition for  ${\cal L}_{01}$ to be a 
globally on $S^2$ defined line bundle is that the 
curvature two-form be in integral cohomology
\beqs
\left[ \frac{f}{2\pi} \right] \in H^2(S^2,\ZZ)~.
\eeqs
As the factor ${\vol}(S^2)$ in the 
formula for the flux (\ref{fl}) 
depends continuously on the choice of $D^3$, we are 
lead to the quantization conditions
\beqs
m_0 &=& m_1 \\
m_0 Q & \in & \ZZ~.
\eeqs
The first one implies that a D8-brane would separate gerbes from 
each other.
The cosmological constant $m_0$ is quantized \cite{Green:1996bh}
because of the Chern--Simons terms in 
the anomalous couplings. Here this leads to a quantization of 
$Q$ as well.

Without the help of gerbes it would be difficult to see how the 
D6-brane charge
should be quantized, or conserved, as the field strength $f$ is 
not globally 
defined, and as its Chern class can be changed by 
NS gauge transformations. 
Here the quantization follows from imposing the 
gerbe structure explained in Section \ref{gerbe}. 
Conservation follows from 
the fact that the flux arises from comparing two  
$B$-fields; 
even if the $B$-fields are shifted by a non-trivial two-form 
in the cohomology of the three-sphere 
(minus the point)
the flux remains the same, provided $m_1 = m_0$. This is exactly 
the 
condition that we stay inside a given gerbe. 

The quantity that is invariant in field theory is $R^{[2]}$. 
In our example there are no electric sources for this field 
$\dd * R^{[2]}=0$, but
\beqs
\dd R^{[2]} &=& m_0Q~ (V - 2\pi \delta_p)~.
\eeqs
This integrates to zero over the sphere, which just means that 
the sphere is able to absorb the 
flux  emanating from $p$.
From the field theory point of view this is a consequence of 
the quadratic
coupling between $B$ and $\dd C^{[1]}$ that turns flux into 
the B-field 
\cite{Strominger:1996ac, Townsend:1996em}. The 
closest parallel to this is the anomalous coupling between 
the Chan--Paton field 
strength and the RR potentials. There will be a net flux out 
of the sphere only 
if it is placed on top of the D8-brane, where the flux can flow. 

The flux of $G$ emanates from the origin where there 
is a source. Except exactly at the D6-brane this flux turns 
into the {\em charge density} of $R^{[2]}$ implying a 
monopole distribution spread evenly over the sphere. Of course 
once we embed the sphere in $\RR^4$ the density falls off with 
the distance 
from the NS5-branes. The physical sources in string theory 
are the D6-branes with which one associates the locally defined 
beam of flux $f$ through the point $p$.

\section{T-duals in type IIB}

Massive type IIA solutions can be related to type IIB solutions 
using a Scherk--Schwarz compactification, where the compact     
coordinate $z \in S^1$ has period $l_B$. Applying 
T-duality on the D6-brane world-volume of 
the configurations in Sec.~\ref{ansatz}  and using the
massive T-duality  rules of \cite{Bergshoeff:1996ui, 
Bergshoeff:1995as}  yields\footnote{The Hodge star 
refers again to the metric on $S^3$ with volume $2\pi$.}
\beqs
C^{[0]} &=& \chi + m z \\
C^{[2]} &=& -~C^{[0]} Q *\dd h + J^{[2]}~,
\eeqs
where $\chi$ is the vacuum expectation value and 
the current $J^{[2]}$ satisfies
\beqs
\dd J^{[2]} = 2\pi~ Q~C^{[0]}~ \delta_p~. \label{currJ}
\eeqs
The exact forms of $C^{[6]}$ and $C^{[8]}$ depend on the 
(yet unknown) metric. 

Notice in passing that 
restricting to the branes $i: M^{1,5} \longrightarrow X$ 
the pull-backs $i^*[G] =0$ vanish, and any spin$_c$-manifold  
$M^{1,5}$
satisfies the consistency condition 
\beqs
i^*[H] = W + [D] \label{Wcons}
\eeqs
for a brane to wrap on it as it is 
stated in \cite{Witten:1998xy}. Here $H = H_\ns, H_\rr$, $[D]$ 
is the Poincar\'e dual of the (here absent) boundary of a 
three-brane ending 
on the NS5/D5-brane, and $W$ is some characteristic class 
constructed 
naturally out of the topology of the brane. Most notably it could 
be $\beta (w_2)$, where $\beta: H^2(X,\ZZ_2) \longrightarrow 
H^3(X,\ZZ)$ is the Bockstein homomorphism -- or possibly the 
characteristic class of 
the brane, if it happens to be a gerbe cf.~Sec.~\ref{Kteoria}.

\subsection{Type IIB charges}

The conserved  charges are
\beqs
q_7 &=& \int_{S^1} \dd C^{[0]} =ml_B\\
q_{\mathrm{D}5} &=& \int_{S^3} \dd  C^{[0]} 
\wedge Q\dd^\dagger H \label{D5integ} \\
q_{\mathrm{NS}5} &=& \int_{S^3} \ee^{2\phi}G -\half 
\dd  \Big(C^{[0]}\Big)^2 \wedge~ Q \dd^\dagger H~. 
\label{NS5integ}
\eeqs
The eight-form charge  signals the presence of 
D7-brane sources cf.~Sec.~\ref{metric}.
If the bounding sphere $S^3$ here is the same 
as the one in the IIA picture we obtain 
the NS5-branes we started from.

Let us show now that it is possible to choose the 
topology in such a 
way  that a non-zero flux could be captured 
by some other three-surface than the previously used $S^3$. 
We could choose this surface to be a circle 
bundle $\sigma: M^3 \simeq S^3 \longrightarrow S^2$
with the same base manifold $S^2$ we used for the 
cylinder $\pi: C \longrightarrow S^2$. 
The fiber will be the isometry direction $S^1$.
Above, we got a generator of $H^2(S^2, \ZZ)$ by pushing forward
$\dd^\dagger(H-H_1)$ by $\pi$. This is the Poincar\'e dual of 
the fiber 
of the cylinder bundle, and only well defined in $C$.
Let us now choose a section of the cylinder bundle 
$i: S^2 \longrightarrow C$. The form $i^* \dd^\dagger H$ is again 
proportional 
to the generator of $H^2(S^2)$ and carries information about 
the choice 
of fibration.
We can pull this form back to $M^3$ using $\sigma$. 
Assuming that $H^1(M^3)=0$ any closed one-form is exact, and 
$\sigma^* i^* \dd^\dagger H = \dd (\omega_h + \omega_v)$ 
where $\omega_v$ 
is an arbitrary closed
vertical form. 
There is indeed a field that we should associate with the vertical form, 
namely $\omega_v = \dd C^{[0]}$. 
These choices of the components of $\omega_{h,v}$ carry information 
about how the isometry 
direction $S^1 \in z$ is embedded in $M^3$. 
The D5-brane charge turns out to 
be related to the linking number (Hopf invariant) 
\beqs
{\cal H}(\sigma) &=& \int_{M^3} \omega \wedge \dd \omega \\ &=&   
\frac{4\pi}{Q~\vol(S^2 \subset S^3)} \int_{M^3} \dd C^{[0]} \wedge B~,
\eeqs
where $\omega = \omega_h + \omega_v$.

The charges obtained from integrating over the
two different spheres span a sublattice 
\beqs
Q~\mathbf{q}   ~+~ 
Q{\cal H}(\sigma)~\mathbf{p}
\eeqs
in the space of NS5/D5-brane charges. The basis here is given 
in terms of the moduli $\phi$ and $\chi$ as
\beqs
~\mathbf{q} &=& \Big(0,~ 2\pi \ee^{2\phi} \Big) \\
~\mathbf{p} &=& \frac{1}{4\pi}~ \vol(S^2)~ \Big(1,~ -2\chi \Big)~, 
\label{dyon2}
\eeqs
where the ordered pair on the right-hand side refers to the charges 
in (\ref{D5integ}) -- (\ref{NS5integ}). 
Periodicity of $z \in S^1$ makes the modulus $
\chi$ periodic as well.

\subsection{Topology of the ansatz}
\label{topo}

Let us recapitulate the construction thus far. 
In type IIA the configuration was described in terms of a gerbe. 
We were considering a system that included NS5-branes,  
and locally defined D6-branes that ended in the NS5-branes.  
If there was a compact isometry direction in the world-volume 
of the D6-brane we could T-dualize the system. We considered, 
for simplicity, a dual system with parallel NS5-branes and 
D5-branes. 
It turned out that the D6-brane flux could get entangled 
with the isometry 
direction in such a way as to produce D5-brane charge in 
addition 
to the previously found NS5-brane charge. Hopf invariants classify 
knots on $S^3$ so that, 
in a sense, we have tied the type IIA D6-brane into a knot on the 
type IIB side.

The obtained geometry, though maybe convoluted if 
described as above, 
is not unfamiliar. We may combine the isometry circle 
and the fiber of the 
cylinder $C$ into another cylinder $C'$ -- the transverse 
space has then the structure of 
a cylinder bundle over a sphere, where the cylinder 
$C'$ is the fiber 
and the sphere  is the usual base space.
This space can then be related to  an elliptically over a 
sphere fibered  four-manifold with a singular fiber. 
Furthermore, there is D7-brane charge, 
which implies a deficit angle in the IIB metric. 
This charge is associated to a world-volume 
which is obtained from the D6-brane ${\cal M}(D6)$ 
modding out by  
isometries $S^1$, and adding the sphere that has been 
serving us as a base manifold, or as 
${\cal M}(D7) = ({\cal M}(D6) \times M^3)/S^1$.
In Section \ref{metric} we shall use 
global consistency conditions to patch these local 
descriptions together 
and relate the geometry of the system to an F-theory  
compactification.

\section{Generalizations}
\label{general}

The construction can be naturally generalized for a 
finite set  of extracted points 
\beqs
{\cal P} = \{p_\alpha|\alpha\in {\cal I}\} \subset S^3~,
\eeqs
by just repeating the above procedure for each point separately. 
One then obtains for each 
$\alpha\in {\cal I}$ a pair of sets 
${\cal U}^\alpha_0$ and ${\cal U}^\alpha_1$. 
Let us define 
$B_0^\alpha = Q \dd^\dagger 
H^\alpha$ and $B^\alpha_1= Q \dd^\dagger H^\alpha_1 $ in 
an obvious generalization of Section \ref{ansatz}. 
We now have to 
ensure that the arising line bundles on each two-fold 
intersection obey the condition (\ref{trivbundle}).

The open neighborhoods of the fixed points can be chosen so 
that they do not intersect. 
We shall have to consider the following intersections of 
local charts
\begin{itemize}
\item[a)] For intersections ${\cal U}^\alpha_1 \cap 
{\cal U}^\alpha_0$ nothing new 
happens, and we get the bundles
${\cal L}_{\alpha}(p_\alpha)$.
The Chern class was
\beqs
-m(B_0^\alpha - B_1^\alpha) = Qm~ \omega_{p_\alpha} 
= c_1\Big({\cal L}_{\alpha}(p_\alpha)\Big)~,
\eeqs
where $\omega_p$ stands for the (properly normalized) 
two-form in (\ref{wrap}).
\item[b)] On ${\cal U}^\alpha_0 \cap {\cal U}^\beta_0$  
we call the bundle ${\cal L}_{\alpha\beta}$.
This intersection is  again topologically a cylinder 
(three-sphere minus 2 points), and one finds
\beqs
 -m(B_0^\beta - B_0^\alpha) 
= Qm~ (\omega_{p_\beta} - \omega_{p_\alpha}  )~.
\eeqs
The Chern class is hence the difference of the separate 
Chern classes
\beqs
c_1\Big({\cal L}_{\alpha\beta}\Big) = 
c_1\Big({\cal L}_{\beta}(\beta)\Big) - 
c_1\Big({\cal L}_{\alpha}(\alpha)\Big)~.
\eeqs
\item[c)] The last cases to define are the intersections  
$ {\cal U}^\beta_0 \cap {\cal U}^\alpha_1 $. Now we get
\beqs
 -m(B_1^\alpha - B_0^\beta) = - Qm~ \omega_{p_\beta}  
= c_1\Big({\cal L}_{\beta}(p_\alpha)\Big)~.
\eeqs
\end{itemize}

On the triple intersection ${\cal U}^\alpha_1 
\cap {\cal U}^\alpha_0 \cap  {\cal U}^\beta_0$
we observe
\beqs
c_1({\cal L}_{\alpha}(p_\alpha)) + 
c_1({\cal L}_{\alpha\beta}) + c_1({\cal L}_{\beta}(p_\alpha))
= 0~,
\eeqs
which means that the product bundle
\beqs
{\cal L}_{\alpha}(p_\alpha) \otimes {\cal L}_{\alpha\beta} \otimes 
{\cal L}_{\beta}(p_\alpha)
\eeqs
is indeed trivial.

Irrespective of the number of points 
the characteristic class is 
\beqs
[H] = Q [V] \in H^3(X,\ZZ)~. 
\eeqs
Notice also that base spaces of the line bundles cannot be 
described as simple embeddings into space-time. 
They are entirely properties of the two-fold 
intersections and hence, in a sense, mutually non-local. 
The gerbe manages to describe the topology of the 
space $S^3 \backslash {\cal P}$; 
in our case it is merely a catalog of the various 
non-trivial two-cocycles in 
$H^2(S^3 \backslash {\cal P}, \ZZ)$. This geometric picture 
is similar to that in e.g.~Refs.~\cite{Sen:1997kz, Sen:1998js}.

\subsection{Consistency and F-theory on a Calabi--Yau three-fold}
\label{Fresol}

In addition to verifying the equations of motion, 
which is a local statement of the stability
of the system, we should also consider global 
consistency. In particular,
due to the deficit angle found above it is not 
consistent to consider only one point ${\cal P} = \{ p \}$. 
We should also take care that the system is at rest, which 
amounts to picking the points in the set ${\cal P}$ where the flux 
comes out of the sphere 
in a  symmetric way. It follows that some 
discrete subgroup $\Gamma$ of 
$\mathrm{SO}(4)$ should act as permutations 
in ${\cal P}$. The system will therefore look 
as if the NS5-branes were put on the top of an orbifold singularity 
in the origin -- the gauge theory on its world-volume 
is therefore the one described in \cite{Blum:1997mm, Witten:1998kz}. 
 
Locally in type IIB 
the transverse space will then look for $\Gamma=\ZZ_n$ like a 
Taub--NUT space. 
In e.g.~\cite{Sen:1997zb} the $B$ field was set proportional 
to the self-dual harmonic two form of the Taub--NUT space, 
the deformations of 
which contributed to the number of zero modes that were 
needed in 
completing the dyon spectrum. 
This provides a type IIB 
interpretation for the ubiquitous factor 
$\vol(S^2)$ 
that appeared in various charges. In our construction it 
characterized the choice of the base manifold 
$S^2 \subset S^3$ using 
which we tie the sphere $S^3$ together with the 
compact direction in 
the Scherk--Schwarz compactification.

\subsection{Geometry of the ansatz}
\label{metric}

In Section \ref{topo} we saw that the 
configurations that support the dyonic fivebrane charges 
have an elliptically over a sphere fibered transverse space, 
and that there is D7-brane charge. 
Above we found  that a stable solution should be invariant under  
$\Gamma$. We have now to patch these 
local cylinder bundles together 
in a way that guarantees the consistency of the 
global metric. 
All of this is rather
suggestive of an orbifold limit of a 
K3 compactification. Related 
configurations have been studied 
e.g.~in \cite{Strominger:1996ac, 
Witten:1995zh, Seiberg:1997zk}. 
A smooth type IIB compactification on a K3 is, however, 
not the right answer as it would have twice as much 
supersymmetry as the original NS5/D6-brane system in
type IIA. Instead, we shall somehow have to take into
account the fivebranes 
that have been put in the orbifold 
singularity in order to break the supersymmetry 
properly to
a quarter of the full type IIB supersymmetry. 

Let us consider for a moment the type IIB system without
the fivebranes.
The fact that our system involves an axion that 
depends on the coordinates of the elliptic fiber was
just a consequence of the Scherk--Schwarz reduction. 
On the other hand, 
any Scherk--Schwarz compactification of type IIB on a circle
can be formulated as an F-theory compactification on a torus bundle
over that circle \cite{Hull:1998vy}. 
In particular, an F-theory solution with the same 
supersymmetry, a monodromy around the D7-branes/orientifolds 
that shifts the axion, and the same 
weak coupling orbifold limit of a K3 singularity 
was  already obtained in \cite{Sen:1997gv, Sen:1997kw, 
Sen:1997bp}: 
The compactification there was an 
elliptically over  $\CC P^1 \times \CC P^1$ 
fibered  
Calabi--Yau three-fold that reduced in ten dimensions in weak coupling 
to type IIB theory on 
$\mathrm{K3}\times {M}^{1,5} / (-)^{F_L} \cdot \Omega \cdot \sigma$
in the notation of \cite{Sen:1997bp}. 

Compactification on K3 and orientifolding
by $(-)^{F_L} \cdot \Omega \cdot \sigma$ both break supersymmetry 
by a half, so that in the end we have the right amount of  
it. The fact that the third cohomology group of a K3 vanishes 
reflects the fact that we must 
put the fivebrane dyons 
in the orbifold singularities of the K3. 
Also the third cohomology group
of $\CC P^1 \times \CC P^1$ vanishes. We see 
that the dyon charges
should arise from topological winding numbers 
only on the level of 
the full Calabi--Yau three-fold whose third cohomology is non-trivial. 
We conclude that we have found at least
a candidate for an F-theory compactification 
that in type IIB gives rise to the right RR charges and 
the right NS5-brane charge. We interpret this  as an 
F-theory argument for, even if not a proof of,  
the consistency of the dilaton and the graviton 
equations of motion\footnote{Notice, however, that 
orientifolding imposes 
restrictions on the two-form potentials.}.

\section{K-theory classification of branes}
\label{Kteoria}

It has been suggested in  
\cite{Minasian:1997mm, Witten:1998cd} and further 
elaborated in \cite{Horava:1998jy, 
Garcia-Compean:1998rg, Gukov:1999yn} 
that K-theory of the world-volume of the brane classify 
D-brane charges. The mapping from 
the brane into the pertinent K-theory class \cite{Witten:1998cd} 
uses the normal bundle of the embedded submanifold, which should have
a spin$_c$-structure. A non-trivial $W \in H^3(Z,\ZZ)$ 
obstructs such a structure, and it follows from the 
consistency condition
(\ref{Wcons}) that the 
the pull-back of the NS three-form field strengths
vanish. 

This situation was generalized to the class of configurations 
involving a NS field strength that is purely torsion in 
integral cohomology. 
In this case one was lead to consider the K-theory
of {\em twisted bundles} \cite{Witten:1998cd}:
These are bundles whose  
transition functions do not necessarily satisfy 
(\ref{2-cocycle}) but instead
\beqs
(\delta g)_{\alpha\beta\gamma} = h_{\alpha\beta\gamma} 
\phi_{\alpha\beta\gamma}~,
\eeqs
where $\phi_{\alpha\beta\gamma}$ is the obstruction to 
spin$_c$-structure
and $h_{\alpha\beta\gamma}$ is the lift of $[G]$ from 
$\check H^3(X,\ZZ)$ to $\check H^2(X,\mathrm{U}(1))$.
These objects are trivial gerbes.

\subsection{Branes ending on branes}

What does happen then, when the restriction of $[G]$ does 
not vanish on the brane? This leads one to consider
branes ending on branes \cite{Witten:1998cd}.  
The world-volume theory will then contain couplings of the 
form $C - \dd L$ of Section \ref{tensorfields} 
(cf. \cite{Strominger:1996ac, Townsend:1996em}) 
and the present considerations are applicable.

The consistent treatment of the world-volume theory involves 
gerbes along the lines of the construction in Section 
\ref{ansatz}, and the characteristic class represented 
by the three-form field strength $[G]$ is, again, that 
of the gerbe on the brane. In the present construction 
the identification of the NS three-form field strength as the
characteristic class of the gerbe 
motivates the point of view that the NS three-form 
incorporates all the topological information of the 
system -- thus including a possible obstruction 
to spin-structure. This leads us too look upon the
consistency condition (\ref{Wcons}) as saying that 
the left-hand side $i^*[G] + W$, which characterizes 
the intrinsic topology of the brane 
whose world-volume theory we are looking at must equal the 
right-hand side $[D]$, which are the sources. 
The natural way to write the consistency equation 
in the form ``topology'' equals ``matter'' is then
\beqs
i^*[G] &=& [D]~.
\eeqs
This condition
is the direct generalization of the formula of electrodynamics 
stating that the Chern class of a Maxwell bundle equals the 
class that characterizes the Dirac string
\beqs
[F] &=& [2\pi \delta_{\mathrm{string}}]~.
\eeqs
Conversely, the obstruction for wrapping a 
brane on a cycle should be the characteristic 
class of the gerbe on the cycle.

\subsection{Classification in type IIA}

The type IIA theory presents 
a couple of problems in the direct application of K-theory: 
The transverse space has odd dimension, and the
analysis in \cite{Witten:1998cd} for type IIB 
has no obvious analogue.
Also, the branes to which the cosmological constant 
$m$ should naively couple
are not submanifolds of space-time -- after all 
their {\em world-volume} 
should have dimension $-1$. Nevertheless string theory tells us 
\cite{polchinski} to treat the cosmological constant 
$m=R^{[0]}$ on equal footing with any other RR field strength. 
Finally, on the level of type IIA supergravity 
the two-form flux is not gauge 
invariant and it is only defined locally, 
as we have seen in the previous sections. 
Instead, the classification group of type IIA 
D-brane charges has been 
suggested in \cite{Witten:1998cd} to be $K^1(X)$. In 
\cite{Horava:1998jy} this picture was 
realized by building supersymmetric D-branes 
starting from non-supersymmetric D9-branes.

The appearance of gerbe solutions calls for an extension 
of the classification to something that 
should be the K-theory of gerbes. However, as the
complications on the type IIA side are 
connected with the appearance of the mass parameter, 
in some cases a straightforward cure 
is simply to T-dualize the configurations to type IIB theory, 
where all field strengths couple to geometric objects. This 
is of course only possible in the configurations that posses 
at least one compact isometry direction needed to perform T-duality, 
and will not circumvent the issues that arise in 
general NS two-form backgrounds. 
The features that were in IIA understood as consequences of having a 
cosmological constant then 
turn into characteristics of a Scherk--Schwarz 
compactification with respect to a coordinate whose period 
is related to the type IIA mass parameter. 
Classification of type IIA D-branes wrapping on a submanifold 
$Z$ should be done in terms of the pertinent group on the type 
IIB side, namely $K(Z\times S^1)$, or some modification 
thereof, where $S^1$ is the circle with respect 
to which the duality transformation was performed. Note that 
the circle may lie also on the D-brane world-volume, 
in which case one might be compelled to 
consider $K(Z/ S^1)$. As in \cite{Witten:1998cd} 
these considerations seem to point towards
the only other available independent K-theory group, namely 
$K^1(Z)$.

\section{Summary}

In this article we have investigated some applications of 
gerbes in string theory and supergravity. 
We showed how these inherently nonlocal 
constructions 
appear naturally in theories that involve massive 
tensor fields analogously to
how  fiber bundles appear whenever there is a 
gauge symmetry. We used this insight in 
analyzing the NS5/D6-brane systems that appear 
in massive type IIA supergravity. 
This leads to quantization conditions 
that, in particular, relate the cosmological 
constant to the NS5-brane charge. The 
D8-branes were interpreted as domain-walls between 
gerbes.
We then proceeded to look for the equivalent 
T-dual systems on 
the type IIB side. The dual branes were 
the NS5-brane and D5-brane dyons. 
Because the used massive T-duality involved 
a non-trivial Scherk--Schwarz compactification 
the axion field became non-trivial, and carried D7-brane 
charge related to the IIA
cosmological constant. 
The system was identified as an F-theory 
compactification on an 
elliptically fibered Calabi--Yau three-fold.

These considerations imply 
foremost that gerbes are  a natural 
part of string theory. The 
immediate reason for their appearance on the 
type IIA side  was the zero-form RR field 
strength, i.e.~the cosmological constant. 
The fact that the corresponding string 
theory objects would be $-1$-dimensional
renders a direct
K-theory classification of D-brane charges
difficult 
on the type IIA side. 
A straightforward method to find the
classification groups 
for type IIA is then to look at 
the T-dual objects on the IIB side.
We also considered Witten's consistency conditions for 
brane wrapping and found that 
they are closely related to gerbe structures. Indeed, the 
integral cohomology class of the NS 
three-form should be interpreted as a 
characteristic class of a gerbe.

We are now in position to use gerbe structures to study 
nonlocal phenomena, such as branes on orbifold singularities 
in type IIB. This picture should also prove 
useful in classifying D-brane charges in the presence of 
non-trivial NS two-form background. 
It would also be interesting to see what 
implications these constructions might have on the spectrum 
of M-theory.

\acknowledgments 
I thank G.~Ferretti and R.~Iengo 
for discussions, and  Helsinki Institute of Physics, where
this project was initiated, for hospitality.

\end{document}